# Conflicting Effects of Extreme Nanoconfinement on the Translational and Segmental Motion of Entangled Polymers

*R. Bharath Venkatesh, Daeyeon Lee\**

AUTHOR ADDRESS: Department of Chemical and Biomolecular Engineering, University of Pennsylvania, Philadelphia, Pennsylvania 19104, USA

\* Email: daeyeon@seas.upenn.edu

ABSTRACT

Physically confining polymers into nanoscale pores induces significant changes in their dynamics. Although different results on the effect of confinement on the dynamics of polymers have been reported, changes in the segmental mobility of polymers typically are correlated with changes in their chain mobility due to increased monomeric relaxation times. In this study, we show that translational and segmental dynamics of polymers confined in disordered packings of nanoparticles can exhibit completely opposite behavior. We monitor the capillary rise dynamics of entangled polystyrene (PS) in disordered packings of silica nanoparticles (NPs) of 7 and 27 nm diameter. The effective viscosity of PS in 27 nm $SiO_2$ NP packings, inferred based on the Lucas-Washburn equation, is significantly smaller than the bulk viscosity, and the extent of reduction in the translational motion due to confinement increases with the molecular weight of PS, reaching 4 orders of magnitude reduction for PS with a molecular weight of 4M g/mol. The glass transition temperature of entangled PS in the packings of 27 nm $SiO_2$ NPs, however, increases by 45 K, indicating significant slowdown of segmental motion. Interestingly, confinement of the polymers into packings made of 7 nm $SiO_2$ NPs results in molecular weight-independent effective viscosity. The segmental dynamics of PS in 7 nm $SiO_2$ NP packings are slowed down even further as evidenced by 65 K increase in glass transition temperature. These seemingly disparate effects are explained by the microscopic reptation-like transport controlling the translational motion and the physical confinement affecting the segmental dynamics under extreme nanoconfinement.

**INTRODUCTION**

Infiltration of polymer into nanoscale pores bears significant importance for a number of processes and phenomena.[1–10] Capillary rise of polymer into anodized aluminum oxide (AAO) membrane has been used to prepare polymer nanowires and nanotubes.[11–13] For example, the Rayleigh instability of polymeric fluids inside AAO pores has been used to template polymeric nanorods with unique curvature.[14,15] Nanoporous templates made from poly-(N-isopropyl) acrylamide have been filled with liquid poly(dimethylsiloxane) and then crosslinked to obtain composites of hydrophobic and hydrophilic materials.[16] Capillary rise infiltration (CaRI) of polymers into the interstitial pores of nanoparticle (NP) packings [2,3,5,17–20] has emerged as a powerful approach to prepare nanocomposite films and membranes with extremely high volume fractions of nanomaterials. Such polymer-infiltrated NP films have superb fracture toughness, thermal stability and corrosion resistance properties.[3,17,21,22] In addition to materials fabrications, infiltration of high molecular weight polymers into nanopores is an important step in enabling upcycling of polymers through heterogeneous catalysis.[23]

In many instances polymer chains undergoing capillary rise into nanoporous media are subjected to extreme nanoconfinement as chains are confined to pores that are substantially smaller than their unperturbed chain dimensions. Such nanoconfinement can bring about significant changes to the ability of the polymer chains to undergo translational and segmental motion. For example, studies on the motion of unentangled polystyrene (PS) into silica ($SiO_2$) NP packings reported an increase in the effective viscosity and the glass transition temperature ($T_g$) of the polymer with increase in the extent of confinement.[24,25] Confinement slows down the polymer at the chain level leading to an increase in the effective viscosity by two orders of magnitude for the most heavily confined unentangled polymers. Contrary to reports of slowdown in unentangled polymers under confinement, a decrease in the viscosity of confined polymers has been reported for the capillary rise of high molecular weight (MW) polystyrene (PS) and poly(ethylene oxide) (PEO) into the nanopores of anodized aluminum oxide (AAO) membranes[26–28] as well as in the diffusion studies of entangled polymers in cylindrical pores.[29–31] The $T_g$ of PEO decreased by 6k in 65 nm diameter AAO nanopores compared to that of the bulk PEO, whereas the $T_g$ of PS undergoing capillary rise in AAO nanopores showed little change compared to the bulk values.[26,27] The segmental dynamics of unentangled PS and poly(2-vinyl pyridine) (P2VP) in the CaRI films of silica were found to be significantly reduced due to geometric confinement.[32] In packings of 7 nm silica NPs, for example, the $T_g$ of unentangled PS increases by 32 K indicating huge levels of segmental slowdown arising from nanoconfinement. These results show that the exact nature of changes in the translational and segmental motion of polymer chains depends on the geometry and the extent of confinement. In general, however, the changes in the translational dynamics of polymers are consistent with those observed in segmental motion in the corresponding systems. Indeed polymer dynamics at different length scales are known to be correlated; that is, a reduction in segmental mobility typically corresponds to reduction in chain mobility due to increased monomeric relaxation times.[33–35] To our knowledge, very few reports have shown the opposite trend – slowdown of segmental dynamics, for example, accompanied by speed up of chain motion.

In this study, we investigate the effect of extreme nanoconfinement imposed by disordered packings of $SiO_2$ NPs on the capillary rise dynamics and segmental motion of high molecular weight PS. We track infiltration dynamics of high MW PS into disordered $SiO_2$ NP packings using *in situ* ellipsometry and extract the effective viscosity of the confined polymers using the Lucas-Washburn equation. We also measure the $T_g$ of PS under confinement using ellipsometry by measuring refractive index changes with temperature on a controlled cooling ramp. Our results show that the confinement can have conflicting effects on the imbibition dynamics and segmental motion of the polymers; while the capillary rise dynamics of the polymer is enhanced in 27-nm $SiO_2$ NP packings, their segmental motion dynamics is reduced. Despite the segmental slowdown, the chains show faster than bulk motion with almost four orders of magnitude reduction in the effective viscosity for the longest chains (PS 4M with unperturbed $R_g$ ~ 50 nm) in the 27 nm NP packings. In 7-nm $SiO_2$ NP packings, translational motion shows little dependence on their molecular weight while their segmental motion is substantially slowed down. These conflicting effects are reconciled by considering the different length scale and time scales of these motions and the effects controlling them. We discuss the possible reasons behind these findings, which could potentially provide important guidance on the fabrication and applications of highly loaded films and membranes prepared *via* CaRI.

**EXPERIMENTAL METHODS**

*Materials*. Polystyrene (PS) of different molecular weights ($M_n$) :173,000 g mol$^{-1}$ (173K) PDI = 1.06, 498,000 g mol$^{-1}$ (498k) PDI = 1.08, 1,000,000 g mol$^{-1}$ (1M) PDI = 1.05, 2,100,000 g mol$^{-1}$ (2.1M) PDI = 1.15, 4,049,000 g mol$^{-1}$ (4M) PDI = 1.09 is purchased from Polymer Source Inc. Aqueous suspension of silica($SiO_2$) nanoparticles (NPs) with average diameters of 7 nm (Ludox SM-30, 30 wt % suspension in water) and 27 nm (Ludox TM-50, 50 wt % suspension in water) are purchased from Sigma-Aldrich. Silicon wafers (10 cm diameter, 0.5 mm thickness, 0-100 Ohm-cm resistivity) are procured from University Wafers.

*Preparation and characterization of polymer-nanoparticle bilayer films*. Silicon wafers are cut into approximately 1 cm × 1 cm squares. The wafers are rinsed with acetone, isopropanol, and deionized water and dried with nitrogen. Subsequently, the wafers are treated by oxygen plasma treatment for 4 minutes. The $SiO_2$ NP suspension is diluted to 10–15 wt % with water. The NP solutions are bath-sonicated for at least 4 h and filtered using a 0.45 μm hydrophilic syringe filter purchased from Fisher Scientific. The solutions are spin-coated on top of the cleaned silicon wafers at 2200–2500 rpm for 1.5 min to get NP packings of 250-300 nm thickness.

To prepare PS films, 2-2.5 wt% PS solution is prepared by dissolving PS in toluene, and then sonicated and filtered using 0.20 μm hydrophobic PTFE syringe filter purchased from Sigma-Aldrich. 200-250 nm thick PS layers are prepared by spin-coating the PS solution at 2000 rpm for 90 s onto cleaned silicon wafers using a WS-400BZ-6NPP/Lite spin-coater from Laurell Technologies Corporation. The polymer film is cut at the edges using a razor blade without damaging the substrate. The film is then immersed slowly into a water bath at an angle of 45° to the water surface such that the PS film delaminates from the substrate and floats on the water surface. The entangled nature of the PS film enables delamination from the substrate. NP film, spin coated earlier on the silicon substrate, is then immersed in the water bath to capture the free-standing polymer film on top of the NP packing to get a polymer-NP bilayer. The bilayer is dried overnight.

Scanning electron microscopy (SEM) images of the cross-section of the samples are taken using a JEOL 7500F HRSEM. Cross-section images are taken by cleaving the sample using a diamond scribe and mounting the sample vertically on a stub with the cross section facing the electron beam. The samples are sputtered with a 4 nm iridium layer using a Quorom plasma generating sputter coater prior to imaging to prevent charging. An accelerating voltage of 5 kV and emission current 20 μA at a working distance of ~8 mm are used to image the samples.

*Characterization of polymer capillary rise infiltration (CaRI) process*. Polymer infiltration into the voids of the NP packing is monitored *in situ* using a J.A. Woollam Alpha-SE spectroscopic ellipsometer. A bilayer sample is annealed above the glass transition temperature ($T_g$) of the polymer using a Linkam THMS350 V heating stage mounted on the ellipsometer. A temperature ramp is specified using the Linksys software which increases the temperature of the heating stage at the rate of 30 °C/min to the desired set point temperatures (155, 160, and 165 °C) and holds it for a specified time (10-12 hours) at the set point temperature. The sample is adhered to the heating stage using a thermal paste Arctic Silver Ceramic polysynthetic thermal compound which allows good contact without thermal insulation. To prevent heat loss due to exposure to the atmosphere, the setup is covered by a thermocol ice box purchased from Amazon. Annealing is stopped when the dynamic data stops changing which takes typically 3-12 hours depending on the molecular weight and temperature.

The ellipsometry data are collected in the wavelength ($\lambda$) range of 380–900 nm at an incident angle of 70°. The CompleteEASE software package provided by J.A. Woollam uses a model-based approach for the fitting of the amplitude change ($\psi$) and phase change ($\Delta$) of the polarized light to a three-layer Cauchy model (NP packing, composite, and polymer) on a silicon substrate with a native oxide layer. The Cauchy model for each layer is expressed as $n(\lambda) = A + B/\lambda^2 + C/\lambda^4$ and $k(\lambda) = 0$, where $A$, $B$, and $C$ are the optical constants, $\lambda$ is the wavelength (in μm), and $n$ and $k$ are the index of refraction and extinction coefficient, respectively. The fitting procedure allows extraction of physical parameters - the thickness and refractive index of each layer of the sample.

*Glass transition temperature ($T_g$) measurements.* The glass transition temperature ($T_g$) of the residual polymer and the polymer confined in the NP packing is measured using a J.A. Woollam M-2000V spectroscopic ellipsometer. The post-infiltration sample i.e. the polymer-infiltrated film with a residual polymer layer on top is mounted onto a Linkam THMS 600 temperature-controlled stage attached to the ellipsometer. The *in situ* ellipsometry sampling rate is 1 s and three heating and cooling cycles between 30 and 200 °C are performed for each sample, with heating rate of 30 K/min and cooling rate of 10 K/min, respectively. The thickness and refractive index of the residual polymer film and the composite are determined by fitting the cooling ramp raw data to the Cauchy model. The $T_g$ of the confined polymer for the composite is determined *via* the intersection of the linear fits to the melt and glassy regimes in the plots of nanocomposite refractive index versus temperature. $T_g$ of non-confined PS is determined using the same protocol on the residual polymer layer's thickness versus temperature plot (see Supporting Information for details). Samples had at least 100 nm thick residual polymer layers on top such that the measured $T_g$ in the thin films would be close to the bulk value.

**RESULTS AND DISCUSSION**

We take advantage of the high surface energy of $SiO_2$ NPs and the high curvature of the nanopores in the NP packing to induce capillary rise infiltration (CaRI) of high molecular weight entangled PS into the pores of the NP packings. A bilayer of PS and $SiO_2$ NP packing is prepared by first spin coating $SiO_2$ NPs onto a silicon (Si) wafer and subsequently depositing a free-standing film of PS atop the NP layer. The scanning electron microscopy(SEM) image of the bilayer before annealing as seen in Figure 1(a) shows the randomly distributed, interconnected, nanometer-sized pores in the nanoparticle packing with the PS layer on top of the packing. Heating this bilayer above the glass transition temperature ($T_g$) of PS (~100 °C) leads to the imbibition of PS into the pores of the packing *via* capillarity. The infiltration process is monitored by spectroscopic ellipsometry which allows us to measure the dynamics of the capillary rise of polymers into the packing. PS infiltrates the NP packing uniformly without creating any cracks. A residual PS layer on top of the PS-NP composite layer can be clearly observed as shown in Figure 1(b). Such a sample is subsequently used to characterize the $T_g$ of the confined PS and the unconfined PS in the residual layer.

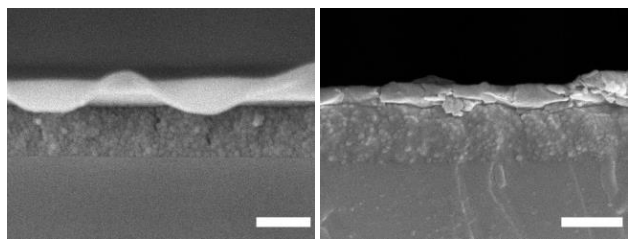

**Figure 1**: (a) Scanning electron microscopy(SEM) Image of PS-Silica bilayer before annealing: the polymer(Polystyrene – 173Kg/mol) film is atop a 250 nm nanoparticle packing made from 30 nm diameter silica nanoparticles and b)SEM image of annealed CaRI film with residual polymer layer on top – the composite layer shows the pores between the particles filled with polymers. Scale bars represent 250 nm.

Prior to the infiltration of PS into the NP packing, the thicknesses of the NP packing and PS layer are determined by fitting the ellipsometry data to a two-layer Cauchy model. The refractive index of the PS layer matches that of a neat PS film. The refractive index for the NP packing is used along with the bulk values of silica and air to infer the porosity of the packing (~ 34%). Upon heating, PS invades the NP packing with a uniform front which we track using a 3-layer Cauchy model composed of a neat polymer layer on top, a composite layer in the middle, and an unfilled NP packing at the bottom. A sharp invading front between the composite and the unfilled NP packing can be detected based on modelling of the ellipsometry data. The thickness of the composite layer gives the height of the infiltrating front with time. PS infiltrates the nanoparticle packing more rapidly at higher temperatures as shown in Figure 2(a).

To quantify the dynamics of capillary rise infiltration, the effective viscosity of PS infiltrating the NP packing is inferred based on the Lucas-Washburn equation: [36,37]

$$h^2 = \left(\frac{\gamma R_{pore} \cos\theta}{4\eta_{conf}\tau^2}\right)t \qquad (1)$$

where γ is the surface tension of the polymer, $R_{pore}$ is the pore radius, θ is the equilibrium contact angle at the triple line between the surface of the pore (silica NP), air and the polymer, $\eta_{conf}$ is the effective viscosity of the polymer melt undergoing capillary rise and τ is the tortuosity of the packing. The average pore size in a random close packing of spheres has been reported to be ~ 30% of the radius of the particle, which we use for $R_{pore}$.[38] The tortuosity of the packing is reported to be 1.95 for interconnected pores in a packing made of randomly close-packed spheres.[39] The surface tension of PS is taken from the bulk values from literature and the contact angle of PS on silica is taken as 20°.[40] The effective viscosity of the confined PS ($\eta_{conf}$) can be calculated based on these values by plotting the infiltration kinetics data as $h^2$ vs t and extracting the slope. The calculated viscosity is the effective viscosity under confinement and it is compared with the bulk viscosity ($\eta_0$) of PS from the published data (see Supporting Information for more details).[41,42]

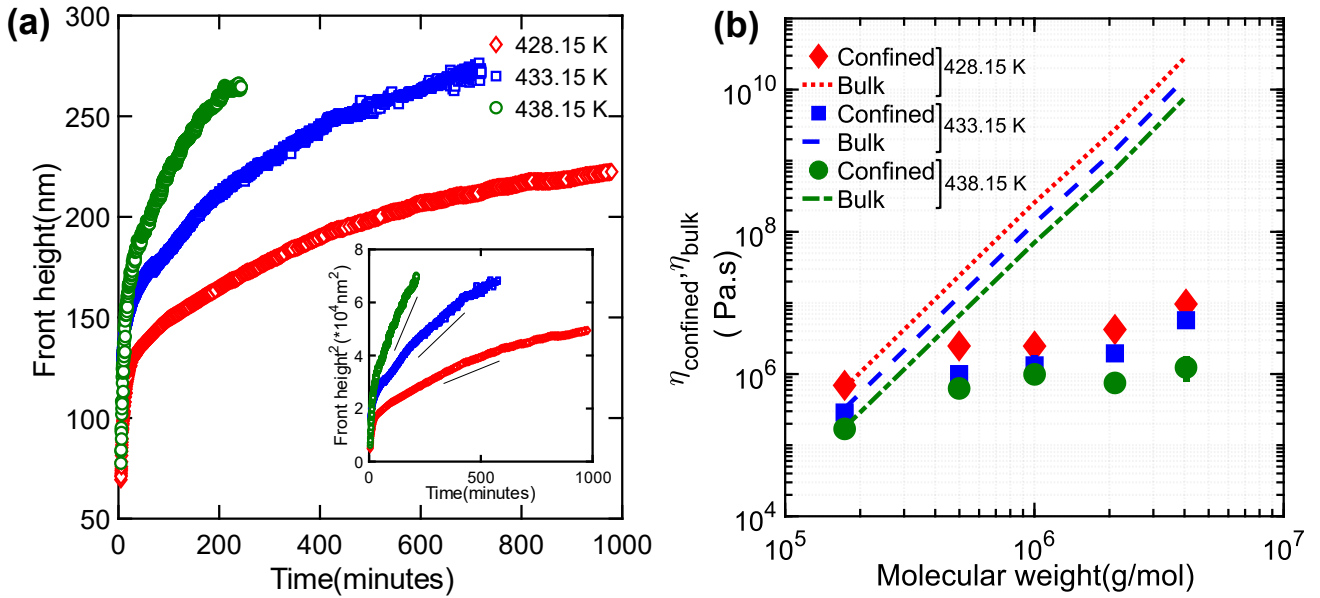

**Figure 2**: a) Front height for PS-4M g/mol infiltrating into silica (27 nm) nanoparticle packings is plotted against the time (in minutes) at three different temperatures – 155 (diamonds), 160 (squares), and 165 (circles) °C. The inset shows the Lucas-Washburn scaling of the dynamic data

when plotted as $h^2$ vs t. The dashed lines parallel to the main curves are indicative of the slope of the curves used to calculate viscosity from the dynamic data. b) Confined viscosity of high molecular weight polystyrene (173k to 4M g/mol) infiltrating silica (27 nm) nanoparticle packings measured using ellipsometric tracking at three different temperatures – 155 (diamonds), 160 (squares), and 165 (circles) °C. The dashed lines show the bulk viscosity calculated using published predictions for zero shear viscosity of entangled PS.

The confined viscosity of PS in 27 nm $SiO_2$ NP packing as well as the bulk viscosity of PS are plotted for three different temperatures as shown in Figure 2(b). The confined viscosity is lower than the bulk viscosity in all cases except for PS-173k at which the two values are very similar. The difference between the two viscosity values increases with molecular weight, and the effective viscosity under confinement is four orders of magnitude lower than the bulk viscosity at MW = 4M g/mol. The effect of MW on the confined viscosity is less significant than that on the bulk viscosity which follows the bulk reptation predictions ($\eta \sim MW^{3.4}$).

The effective viscosity of PS confined in packings of 7 nm $SiO_2$ NPs shows a unique trend quite different from the effective viscosity of PS in 27 nm $SiO_2$ NP packings. As can be seen in Figure 3, the effective viscosity stays largely unaffected across a large range of MW and depends only on the temperature. While the PS-4M shows enhanced translational mobility under confinement compared to the bulk PS, PS-173k has effective confined viscosity that is higher than that of bulk PS. Such an observation, to our best knowledge, has not been previously reported and we offer our hypotheses responsible for this trend later.

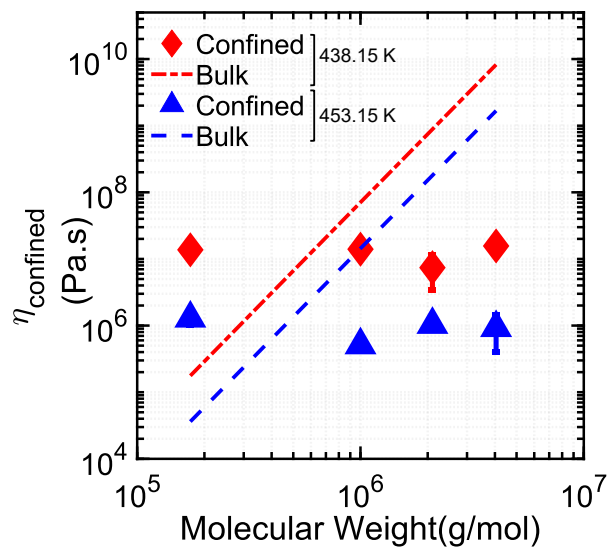

**Figure 3**: Confined viscosity of high molecular weight PS (173k - 4M g/mol) infiltrating into $SiO_2$ (7 nm) NP packings measured using ellipsometric front tracking at two different temperatures - 165 °C (diamonds) and 180 °C (triangles). The dashed lines are the bulk viscosity for high molecular weight PS, estimated based on published results.

---

To check if the enhanced translational mobility of the confined polymers is affected by changes in the segmental motion, the $T_g$ of confined PS in both 7 nm and 27 nm packings are measured and compared to the bulk $T_g$ measured using the residual polymer film atop the composite layer (see Supporting Information for details). Surprisingly, there is a slowdown in segmental motion as seen from the increase in the confined $T_g$ relative to the bulk $T_g$ as shown in Figure 4. The bulk $T_g$ values are close to the previously reported $T_g$ for bulk PS (100-103 °C). In contrast, $T_g$ of PS confined in silica nanopores is about 45 K higher than the bulk PS $T_g$ and does not change much with MW. For PS confined in the pores of 7 nm $SiO_2$ NPs, the difference between the confined and the bulk $T_g$ is even larger (around 65 K) and does not depend strongly on the MW. These observations are consistent with recent reports that showed that PS confined in the $SiO_2$ NP packings has significantly higher $T_g$ compared to bulk,[25,32] and more importantly suggest that the translational motion of high molecular weight PS in the SiO$_2$ NP packings is not closely coupled to their segmental motion.

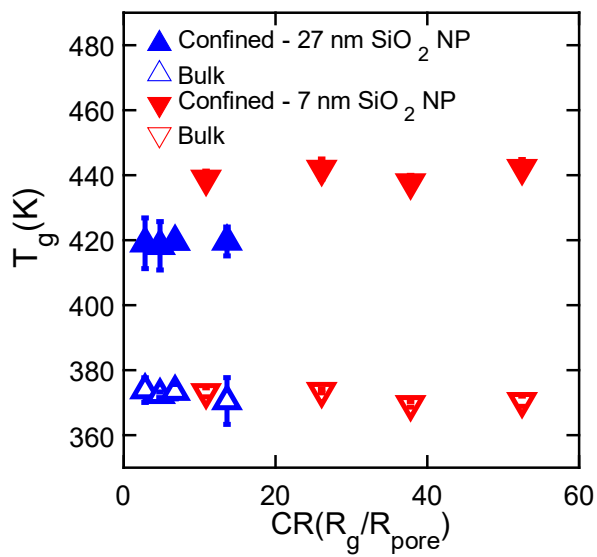

**Figure 4**: Glass transition temperatures (T$_g$) for confined and bulk high molecular weight polystyrene at different confinement ratios in 27 nm (star) and 7 nm (inverted triangle) silica nanoparticle packings. The closed symbols are T$_g$ measurements done for polymer confined within the nanoparticle packings after infiltration and the open symbols are for polymers in the thin residual film (>100 nm) left atop the nanoparticle packing after infiltration. The latter measurements reflect the bulk Tg values of entangled polymers

Polymers are subjected to extremely high degree of confinement (pores of 1-4 nm) in the pores of nanoparticle packings. For example, by varying the molecular weight of PS between 173,000 g/mol(PS-173k) and 4,000,000 g/mol (PS-4M), and using either 27 nm or 7 nm SiO$_2$ NPs, the extent of confinement, defined by the ratio of the radius of gyration of polymer and the average pore size, can be varied between 2 and 60. Thus, at the highest level of confinement, the chain dimension can be an order of magnitude larger than the confining pore dimension. The effect of such extreme nanoconfinement is complex as discussed here.

*Enhanced translational mobility in 27 nm silica NP packings.* To assess how the extent of confinement is affecting the dynamics of confined PS in the $SiO_2$ NP packings, the effective viscosity of PS confined in 27 nm $SiO_2$ NP packings is normalized with respect to the bulk viscosity and plotted as a function of the confinement ratio (CR), which is defined as the ratio of the radius of gyration of PS and the average pore size ($R_g/R_{pore}$). The reduction in the effective viscosity with increasing confinement is clearly seen in Figure 5 with a temperature-independent scaling of the normalized viscosity with CR. Remarkably, at the highest level of confinement (PS-4M, CR ~15), the effective viscosity falls by almost four orders of magnitude with respect to the bulk. Note that the viscosity is normalized with respect to the bulk viscosity at the same temperature and not at a fixed distance from $T_g$. The magnitude of the relative viscosity change is even greater if we take the latter approach but the scaling with CR remains the same (see Supporting Information for scaling with fixed distance from $T_g$).

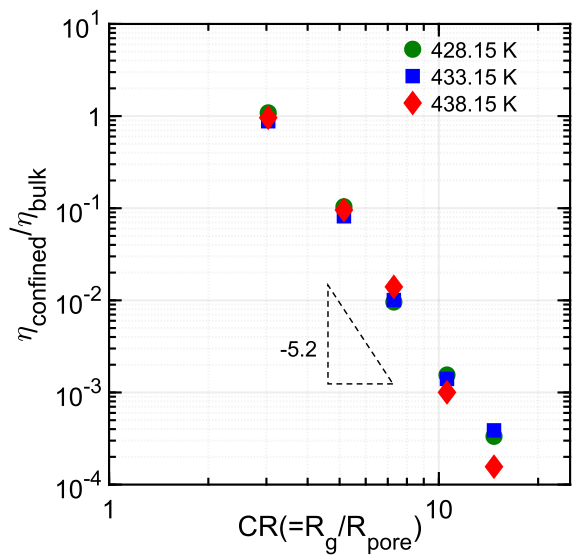

**Figure 5**: Confined viscosity of high molecular weight polystyrene (173 k – 4 M g/mol) infiltrating silica (27 nm) nanoparticle packings is normalized with the bulk viscosity at three different temperatures – 155 (diamonds), 160 (squares), and 165 (circles) °C - and plotted against confinement ratio (CR). The scaling is extracted from the slope of the best linear fit of the data points.

We consider some possible reasons behind the observed reduction in the effective viscosity of PS in 27 nm $SiO_2$ NP packings. MD simulations have shown that dynamic contact angle correction is necessary to use the Lucas-Washburn equation to model the flow of non-Newtonian fluids like polymer melts in micro and nanochannels.[43–45] The measured viscosity, however, has been found to be not sensitive to the dynamic contact angle in a prior study on the capillary rise of poly(ethylene oxide) (PEO) (MW = 100 kg/mol to 1,000,000 g/mol) in the pores (pore diameters = 25 to 400 nm) of anodized aluminum oxide (AAO) membranes.[26] The marginal change in the viscosity by applying the dynamic contact angle correction could not explain the drastic reduction in the viscosity observed in their study. Another possibility is that entangled polymers are known to undergo shear thinning due to chain disentanglement when they are subjected to high shear.[46] The local shear rate in the polymer during capillary rise in the NP packings, however, is not high enough to induce the drastic reduction in the effective viscosity that we observe (see Supporting Information for details).

The enhancement in translational mobility that we observe in this work is similar to the reduction in the effective viscosity measured in prior reports on the capillary rise of PS and PEO.[26,27] A novel mechanism of fluid transport which involves plug flow of entangled polymeric melt in narrow capillaries with no-slip boundary conditions has been invoked to explain the reduction in viscosity.[47] In this mechanism, polymer chains reptate through an entangled mesh of constraints which itself moves under the action of the external pressure gradient, provided by the capillary pressure.[47,48] According to this model, the effective viscosity of entangled polymers during capillary rise can be expressed as:[48]

$$\frac{\eta}{\eta_o} = \left[\left(\frac{R_{eff}}{R}\right)^4 + \frac{\varphi \eta_o}{NR^2}\right]^{-1} \quad (2)$$

where the first term on the right-hand side of the equation accounts for the reduction in chain mobility due to reduction of the pore radius from R to $R_{eff}$ induced by the chains adsorbed on the wall; the second term represents microscopic reptation where φ accounts for the monomeric friction, entanglement length, and other constants. The first term dominates when $R_g \ll R$ in the case of underconfined polymers, whereas the microscopic reptation becomes important for confined polymers where $R_g \gg R$ such as the infiltration of entangled PS into nanopores of $SiO_2$ NP packings studied in our current work. If Eq 2 is simplified by neglecting the first term as is valid for the case of extreme confinement in the nanopores used in this study, $\eta \sim N$.

The normalized viscosity of confined PS in 27 nm $SiO_2$ NP packing with respect to CR shows $\frac{\eta}{\eta_o} \sim \left(\frac{R_g}{R_{pore}}\right)^{-5.199} \sim N^{-2.6}$. When combined with the scaling of zero-shear viscosity of entangled melts ($\eta_o \sim N^{3.4}$), the scaling for confined viscosity can be expressed as $\eta \sim N^{0.8}$ which is close to that expected from the model. This result is also close to the previously reported values of scaling exponents, 1.4 and 0.9 in studies on the capillary rise of PS and PEO into AAO nanopores, respectively.[26,27] The enhancement in mobility of confined PS in AAO nanopores has been explained with the help of disentanglement of chains and reduction in the density of confined chains. This leads to reduction in confined polymer viscosity and an expected scaling of flux with chain size as $N^{1.4}$.[27] The scaling is also similar to the entropic barrier model scaling proposed by Muthukumar[49] using a model developed for highly heterogenous confinement in gels and concentrated solutions; however entropic barriers to polymer diffusion cannot explain the enhanced chain dynamics seen here.

*Molecular-weight independent viscosity in 7 nm silica packings.* High molecular weight polymers due to their high viscosity can exhibit slip at boundaries. When the slip length is high, there is very little radial variation in the velocity profile inside the pores. This constant velocity plug flow due to high slip (where the slip length is so high that it wipes out any dependence of flow on polymer dynamic properties) can be postulated as a reason for the molecular-weight independent effective viscosity of PS in 7 nm $SiO_2$ NP packings. The infiltration dynamics is controlled by the increased viscous resistance coming from the slip length term. If this term is large enough to overpower changes in the intrinsic viscosity with increasing molecular weight, it is possible that the infiltration dynamics exhibit molecular weight independence and thus MW-independent confined viscosity.[50]

Prior studies have reported slippage of wetting fluids on solid surfaces.[51] Considering the favorable interactions between the polymer and the NP (contact angle of PS on $SiO_2$ is 20°), however, it seems unlikely for the polymer to slip on the NP surface. Additionally, the chains adsorbed on the solid walls may entangle with the flowing chains leading to strong suppression of slip. At high shear velocities of the order of μm/s, the adsorbed chains can disentangle from the flowing chains leading to revival of slip.[52,53] As mentioned earlier, however, the shear rate in the CaRI process is not high enough to lead to any significant shear-induced disentangling between surface adsorbed and bulk chains which means that slip is not significant in describing the flow dynamics.

While the reptation-driven plug-flow mechanism can explain the enhanced mobility of confined polymers in 27 nm $SiO_2$ packings and accounts for the linear scaling of confined viscosity with molecular weight, the model cannot explain the constant effective viscosity of confined PS in the packings of 7 nm $SiO_2$ particles. One possible reason is that the model was developed to consider a linear flow regime in which polymers obey Gaussian chain statistics.[47,54] This model would not be accurate under a high capillary pressure generated in such narrow pores where the chain conformation may be substantially perturbed. A theoretical framework for molecular-weight-independent viscosity has been proposed for polymer melts flowing under high pressures.[47] The model states that polymer flowing under high capillary pressure experience increased stretching due to incomplete screening of the excluded volume. This stretching leads to transitioning from the linear flow regime to a non-linear flow regime at a crossover pressure gradient, which depends on the chain size and the pore radius. The reptation model applied to such a non-ideal flow case is shown to behave independently of molecular weight of the polymer (see Supporting Information for details). The high capillary pressures in 7 nm $SiO_2$ NP packings is indeed above such a crossover pressure gradient.

The average pore size in the 7 nm $SiO_2$ packings is 2 nm, which is comparable to the Kuhn segment length of the polymer. Such a state of extreme nanoconfinement is referred to as the hyperconfined condition.[31,54] Polymer chains experience excluded volume interactions due to incomplete shielding under hyperconfinement which leads to extreme axial stretching of the chains.[31,54] To avoid the entropic penalty of stretching, the chains segregate to occupy different regions forming a train of chains occupying pores. Such hyperconfinement occurs above a critical chain size (or MW of polymer) relative to the pore size, and this threshold for the 7 nm $SiO_2$ packings is at a lower molecular weight than for the 27 nm $SiO_2$ NP packings (see Supporting Information for detailed calculations). The range of MW of polymers confined in the 7 nm $SiO_2$ packings are all in the hyperconfined regime. Although we do not fully understand the exact correlation between hyperconfinement and the observed flow behavior, we point out that the level of confinement that leads to the chain size-independent viscosity coincides with the onset of hyperconfinement.

*Segmental slowdown.* Segmental slowdown has been reported previously near attractive surfaces due to the adsorption of chain segments and/or non-bulk like conformation of the polymer chains under confinement. In addition, segmental slowdown has been reported near impenetrable surfaces and highly confined spaces due to reduced free volume available for polymer motion. In the CaRI system, there is significant confinement of the polymers in the pores of NP along with a large interfacial area between the polymer and the NPs. In a previous study on PS confined in the interstices of $SiO_2$ NP packings, a dramatic increase in $T_g$ was observed with increasing confinement (decreasing NP radius) and the effects were mostly understood as originating from the loss of conformational entropy due to geometric confinement of the polymer.[32] Studies on undersaturated CaRI films showed that the polymer exists in the highly confined regions of the packings between the particles leading to high $T_g$.[3,32] Thus, the source of increase in $T_g$ of entangled polymer chains in the 7 nm and 27 nm $SiO_2$ NP packings comes mainly from the high geometrical confinement experienced by the polymer chain in the nanopores. The unique nature of the concave pores also plays a role by decreasing the free volume and increasing the packing density which is strikingly different from PS trapped in convex pores like those in AAO. The pores of the NP packing where the concave nature of the pores can be hindering segmental relaxation, lead to an increase in $T_g$.[55] An additional increase in slowdown of entangled polymers in 7 nm packings over 27 nm packings comes from not only increase in confinement but also increase in the interfacial area between polymer and $SiO_2$ NP. As shown in Figure 4, in the range CR 10-15, the increase in $T_g$ for PS-4M in 27 nm $SiO_2$ NP packings is 50 K, whereas PS-173k in 7 nm $SiO_2$ NP, despite being at the same level of confinement (same CR), shows an increase of 65 K in $T_g$, which indicates that the interfacial area also contributes to the segmental slowdown of polymers in CaRI films.

To check whether an increase in the interfacial area alone can explain the trends in $T_g$, we plot the increase in $T_g$ for unentangled and entangled PS in $SiO_2$ as a function of the interfacial area/volume of the NP packings (see Supporting Information for details). While the large interfacial area is an important factor contributing to the large increase that we observe, it cannot completely account for the trends alone. A combination of increasing confinement and interfacial area between the polymer-nanoparticle can explain the observed $T_g$ trends. As the polymer size is increased (and/or NP size is reduced), the polymer chains experience larger confinement as well as an increase in interfacial area between polymer-NP increases leading to the observed increase in $T_g$.

*Decoupling of translational and segmental mobility.* Unlike earlier reports of PS chains in the pores of AAO membranes which showed the enhanced translational mobility and either little change or an increase in the segmental dynamics,[26,27] the increase in translational motion of confined polymers in $SiO_2$ NP packings accompanies a drastic reduction in segmental mobility; that is, changes in the segmental dynamics do not translate to corresponding changes in the infiltration dynamics. Our results demonstrate that extreme nanoconfinement of polymers can decouple dynamics occurring over different length scales over different time scales. The time-temperature superposition relationship connecting the viscosity and $T_g$ has been shown to fail in some nano-confined systems.[56,57] One notable study, for example, has shown both the $T_g$ and the diffusivity of poly(isobutyl methacrylate) confined between two weakly interacting surfaces increases.[56] An increase in the diffusivity of confined chains was attributed to the reduction in the friction coefficient caused by weakly-interacting interface, whereas the slowdown of segmental relaxation was thought to be affected by confinement.

The conflicting effects of confinement on chain and segmental motion in this work arise from the disparate effects acting independently on different length scales. With increasing confinement, entangled chains participate in a reptation-driven flow into nanopores which leads to faster-than-bulk flow rates. On the other hand, confining polymers in nanoscopic pores with convex curvature and increasing particle-polymer interface slows down segmental dynamics of the confined polymers in the nanopores. Thus, the effects of confinement can be quite different at different length scales for entangled polymers confined in nanoparticle packings. In the next section, we summarize our results on unentangled and entangled polymers to provide a picture of the effect of confinement on the motion of polymers at the chain and segmental level in CaRI packings.

*Summary of the confinement effect on the translational and segmental dynamics of unentangled and entangled polymers in nanoparticle packings.* By combining data on $T_g$ and effective viscosity obtained from the CaRI of unentangled polymers from our prior work[25] with the data on entangled polymers from this study, we summarize the effect of nanoconfinement on the segmental and translational motion of confined polymers over a wide range of confinement ratios and molecular weights. Figure 6(a) shows normalized viscosity of both unentangled[25] and entangled polymers in $SiO_2$ NP packings. While the viscosity of unentangled polymers increases with increasing confinement, the effective viscosity decreases with increasing confinement in CaRI of entangled polymers in 27 nm packings. This can be understood by considering the role of the two terms in Equation 2. The first term is effective for underconfined polymers ($R_{pore}>R_g$) while the second term controls the motion of confined polymers ($R_{pore}<R_g$). In the first term, $R_{eff}$ decreases with increasing confinement which explains the increasing viscosity upon confinement for unentangled

polymers whose CR<1. The second term, which is effective for confined entangled polymers (CR>1), increases with decreasing pore size leading to an effective reduction in viscosity. The motion of entangled polymers occurs by reptation in a mesh of constraints under the action of capillary pressure – this motion is faster than their bulk counterparts. Unentangled polymers are slowed down by the reduction in the pore radius due to the adsorption of polymers on the pore surface. The microscopic flow model developed in Reference 47 qualitatively explain trends in our confined viscosity qualitatively over a wide range of confinement (CR from 1–10). The confinement-independent viscosity of PS in 7nm SiO2 packings, however, requires further consideration and theoretical development.

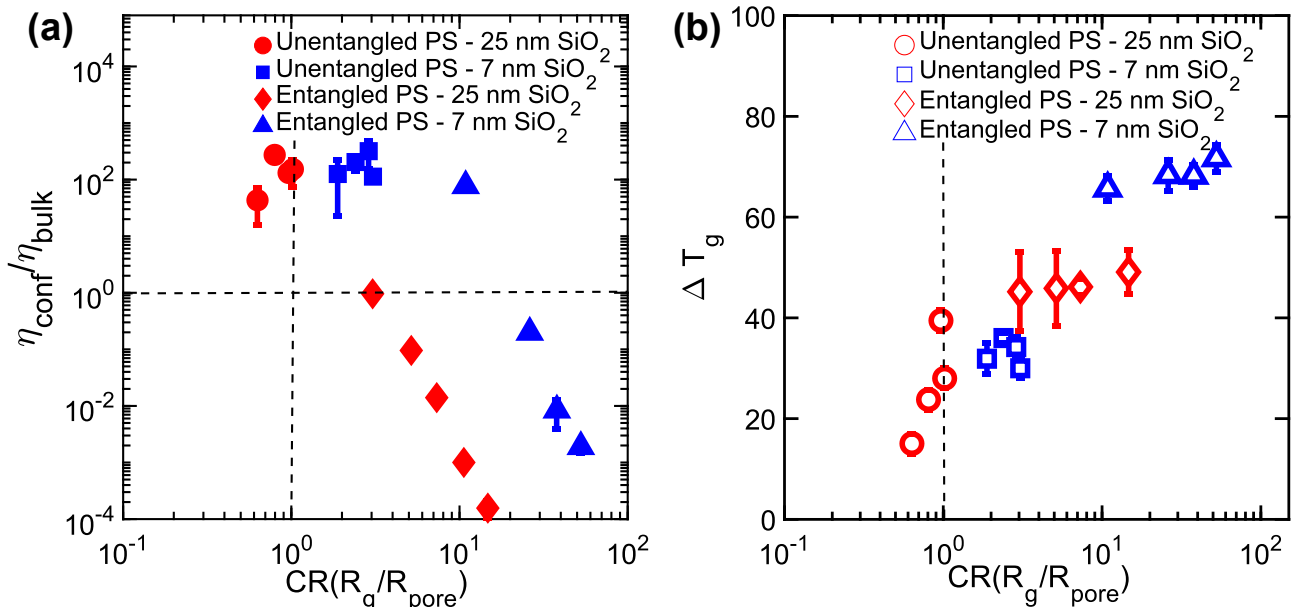

**Figure 6**: (a) Confined viscosity measured for polystyrene (8k g/mol to 4 M g/mol) infiltrating silica nanoparticle packings at different confinement ratios - unentangled PS (8k g/mol to 21k g/mol) in silica of 9-77 nm diameter measured at 130 °C, shown in circles; entangled PS (173k g/mol to 4 M g/mol) in silica of 27 nm measured at 165 °C, shown in diamonds; and entangled PS(173k g/mol to 4 M g/mol) in silica of 7 nm measured at 165 °C, shown in triangles. (b)

Difference between the glass transition temperature in confinement over the bulk ($\Delta T_g$) at various confinement ratios for the three scenarios in (a). Reprinted with permission from the publishers.[25]

Figure 6(b) shows $T_g$s' obtained for unentangled and entangled polymers with increasing confinement. We see that with increasing confinement (smaller pores or larger polymer), the segmental dynamics are further slowed down. This can be understood by combined effect of geometric confinement and the dynamical slowdown of the interfacial layer. In the weakly confined regime (CR<1), the $\Delta T_g$ increases with increasing CR due to increasing confinement. When the chain size becomes larger than the pore size, a further increase in $\Delta T_g$ is observed. The role of confinement is even more evident in the larger $T_g$ increase in 7 nm $SiO_2$ packings (as compared to the 27 nm packings) where due to the larger geometric confinement and an increase in interfacial area, the $T_g$ increases by 65 K over bulk.

The dependence of translational and segmental dynamics on CR in Figure 6 shows that one universal parameter like CR cannot describe the entire trend (data does not collapse with pore size/chain size as seen in previous studies[48,58]). The effect of nanoconfinement on polymer dynamics arises from a complex interplay of chain entanglement, extent of confinement, the pore curvature, and the absolute size of confining pores. The extent of confinement as defined by CR influences the deviations from bulk dynamics both at the chain and the segmental levels. The radius of the pore is another key parameter; we observe that when $R_{pore}$ approaches the Kuhn length of the polymer, the chain viscosity becomes independent of molecular weight – an observation that cannot be fully explained by the existing models of polymer dynamics.

**CONCLUSIONS**

Chain and segmental dynamics of entangled polystyrene infiltration by CaRI into silica nanoparticle packings are studied using ellipsometry to probe the effect of nanoconfinement on polymer mobility. In the range of confinement ratios (CR from 3 - 55) probed, increasing confinement has opposite effects on the polymer dynamics at the chain and the segmental level. Capillary rise dynamics of entangled PS in 27 nm $SiO_2$ NP packing show that the confined viscosity is reduced when compared to the bulk viscosity. Contrary to enhancement in translational mobility, confinement reduces the segmental motion across all levels of confinement with the effect increasing with decreasing particle size owing to an increase in confinement along with greater interface between the polymer and the attractive $SiO_2$ NP. We also report molecular-weight-independent value of effective viscosity in the 7 nm $SiO_2$ NP packings. The trends in the confined viscosity are explained qualitatively by theoretical arguments of dead-zone of adsorbed polymers slowing down chain motion for underconfined polymers, whereas reptation-like plug flow leading to enhanced microscopic flow of entangled chains. Reduced segmental mobility is understood to be an effect of high confinement of polymer chains in the concave pores of the nanoparticle packing and the large interfacial area between silica and polymer under confinement. Thus, the effect of confinement on motion of polymers is different at different length scales and depends on the relative role of the macroscopic and microscopic mechanisms of motion. The enhanced translational motion of high molecular weight polymers can be used to facilitate their infiltration into nanoparticle packings to fabricate highly loaded nanocomposite membranes and films. Moreover, such enhanced motion of polymers in nanopores is advantageous for performing reactions to enable upcycling of polymers *via* heterogenous catalysis which typically require intimate contact between nanoporous catalysts and polymers. Increasing $T_g$ also could be advantageous for applications that require high temperatures, enabling wider operating conditions for composite materials prepared *via* CaRI.

## ACKNOWLEDGMENTS

We thank Prof. Zahra Fakhraai, Dr. Haonan Wang, and Yueli Chen for their help with the ellipsometry measurements. This work was supported by Penn MRSEC (DMR 1720530). This work was carried out in part at the Singh Center for Nanotechnology, which is supported by the NSF National Nanotechnology Coordinated Infrastructure Program under grant NNCI-2025608.## ASSOCIATED CONTACT

The Supporting Information is available free of charge.

Bulk viscosity, contact angle, and surface tension calculations, Tg extraction from the ellipsometric data, Constant viscosity with MW in 7 nm silica nanoparticle packings, Role of interfacial area of PS-silica on increase in Tg, Viscosity data plotted against 1/r, Shear rate calculations, Scaling of viscosity at fixed distance from Tg.(PDF)

## AUTHOR INFORMATION

**Corresponding author**

Daeyeon Lee - Department of Chemical and Biomolecular Engineering, University of Pennsylvania, Philadelphia, Pennsylvania 19104, USA. Email: daeyeon@seas.upenn.edu

**Authors**

R Bharath Venkatesh - Department of Chemical and Biomolecular Engineering, University of Pennsylvania, Philadelphia, Pennsylvania 19104

**Author Contributions**

The manuscript was written through contributions of all authors. All authors have given approval to the final version of the manuscript.

**Notes**

The authors declare no competing financial interests.

ABBREVIATIONS

CaRI: Capillary Rise Infiltration, NP: Nanoparticle, $SiO_2$: Silica, PS: Polystyrene, $T_g$: Glass transition temperature, CR: Confinement ratio.

Supporting Information for

# Conflicting Effects of Extreme Nanoconfinement on the Translational and Segmental Motion of Entangled Polymers

*R. Bharath Venkatesh, Daeyeon Lee\**

AUTHOR ADDRESS: Department of Chemical and Biomolecular Engineering, University of Pennsylvania, Philadelphia, Pennsylvania 19104, USA

# 1. Bulk viscosity, contact angle, and surface tension calculations

The surface tension of bulk Polystyrene varies with temperature and molecular weight as:[1,2]

$$\gamma = \gamma_\infty - \frac{k}{M^{\frac{2}{3}}} \qquad (S1)$$

where $\gamma_\infty$ varies with temperature(T), k is a constant, and M is the molecular weight. The following expression has been developed for MW = 240,000 g/mol:

$$\gamma = 34.60 - 0.05T(in\ °C)\frac{mN}{m^2} \qquad (S2)$$

for calculating polymer surface tension at the different temperatures. From literature,

$$\gamma_\infty = 42.3 - 0.071 * T(in\ °C)\frac{mN}{m^2} \qquad (S3)$$

and using equations S1-S3 at 155 °C for MW=240,000 g/mol, we find

$$\gamma = 42.3 - 0.071 * T - \frac{17166.401}{M^{\frac{2}{3}}}\ \frac{mN}{m^2} \qquad (S4)$$

Contact angle comes from nanodroplet measurements of polystyrene on a thick silicon oxide layer atop a silicon wafer using AFM in tapping mode,[3] $\Theta = 20°$.

Bulk viscosity is obtained from Fox-Flory relationships[4–6] developed for entangled PS(M>38,000 g/mol):

$$\log \eta_{217} = 3.4 \log(M) - 13.40 \qquad (S5)$$

$$\log\left(\frac{\eta}{\eta_{217}}\right) = 2.68 * 10^{10} * e^{-\frac{1330}{M}} * \left(\frac{1}{T^6} - \frac{1}{490^6}\right) \qquad (S6)$$

where $\eta_{217}$ is the viscosity at T=217 K which calculated and then used to calculate the viscosity at desired molecular weight and temperature. Here, the temperature is in Kelvins.

## 2. $T_g$ extraction from the ellipsometric data

The refractive index of PS in the confined composite layer is measured in a controlled cooling step (at the rate of 10K/min) along with the thickness and refractive index of the residual PS film. Figure S1 shows the change in slope of the measured values with temperature occurring at the glass transition temperature for the bulk and the confined polymer films. Due to capillary condensation of water vapor in the nanoparticle packing at T<100 °C, there is a deviation in the slope of the composite refractive index profile (not shown here). This region was excluded in the linear fitting while calculating $T_g$. There is a single $T_g$ in the measured range with no significant breadth in the temperature range of glass transition.

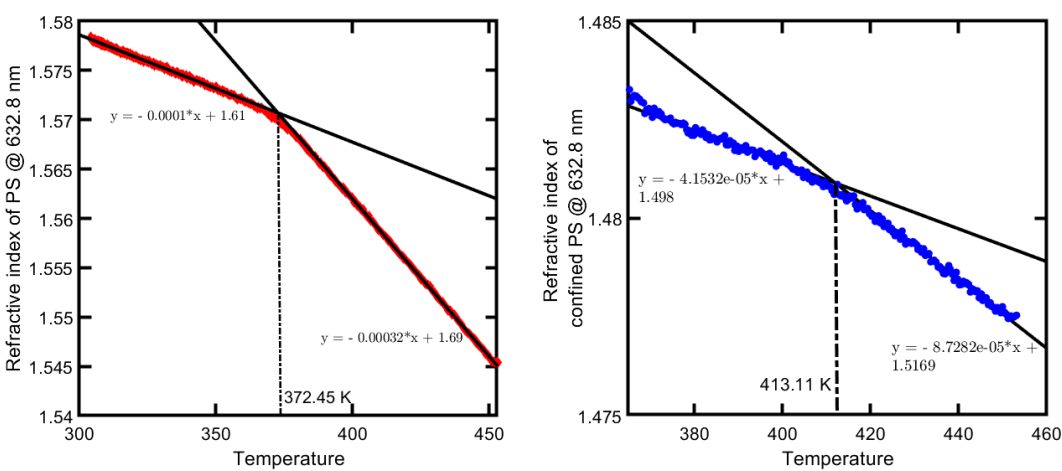

**Figure S1**: Extracting $T_g$ for bulk(left) and confined(right) polystyrene by tracking changes in slope of measured refractive index with temperature in a controlled cooling ramp.

## 3. Constant viscosity with MW in 7 nm silica nanoparticle packings

(a) Non-linear flow arising from a high capillary pressure

The non-linear flow regime occurs when the flow perturbs the chain structure which in turn has impact on the dynamics of the flow. The critical pressure postulated by Johner et al[7] for crossover from the linear to the non-linear flow regime is obtained by matching the energy drop across a stretched chain to the thermal energy. The expression for the pressure drop is given by:

$$p_{co}' = \frac{K_B T}{N_e^{\frac{1}{2}} N b^4} \qquad (S7)$$

where the thermal energy is given by $K_B T$, $N_e$ gives the number of monomers in an entanglement strand, N is the number of monomers, and b is the Kuhn length of the polymer chain. Thus, this crossover pressure depends on the chain length.

The pressure gradient in the nanoparticle packing of particle radius r (and pore size = 0.3r) comes from the capillary pressure drop across the wetting meniscus of the polymer. If the wetting meniscus size can be considered comparable to the pore size, the pressure gradient can be written as:

$$p_{cap}' = \frac{\gamma \cos\theta}{0.09 * r^2} \qquad (S8)$$

The pressure drop across the backbone of the polymer which would lead to non-linear polymer flow based on this model is around 10 KPa (0.1 bar) for PS-173k. Accounting for Laplace pressure difference, the pressure gradient experienced by the rising polymer in the 7 nm particle packings is $1.39 \times 10^7$ KPa (13,800 bars). We also note that the pressure gradient in the 25 nm silica packings is also $4.12 \times 10^5$ KPa (4.12 bars) which is also higher than the critical pressure. Hence, the non-linear flow effects likely is not the only source for the constant viscosity observed in the 7 nm silica particles.(b) Hyperconfinement – onset of hyperconfinement coincides with the constant viscosity seen in 7 nm silica nanoparticle packings

The mean pore size in the 7 nm silica nanoparticle packing is 2.1 nm which is comparable to the Kuhn length of the polymer chain (1.8 nm). This leads to severe nanoconfinement of the polymer giving rise to a hyperconfined regime postulated by several authors.[8,9] The hyperconfined regime leads to complete segregation of chains to avoid the entropic cost of stretching the chain along the flow direction. This argument is similar in spirit to non-linear effects because it considers the perturbation to the Gaussian chain statistics of the flowing chain but the resultant effect on dynamics have not been understood so far. Reports mention that hyperconfined polymers may experience severe slowdown. While we do not fully understand the link between hyperconfinement and constant flow rate, we point out here that the onset of hyerconfinement

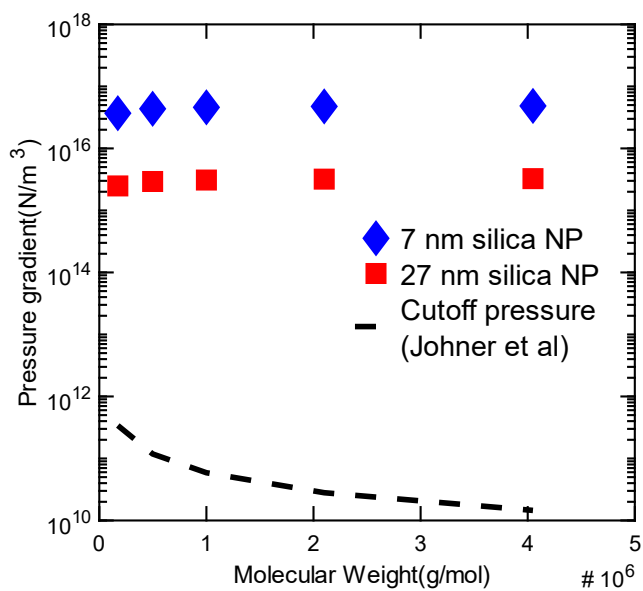

**Figure S2**: Capillary pressure gradient in NP packings of 7 nm (diamonds, blue) and 27 nm (squares, red) SiO$_2$ NPs and critical cutoff pressure postulated by Johner et al. (dashed black line) plotted for varying molecular weight of entangled PS (173k g/mol to 4M g/mol).

condition coincides with the smaller pore radius of the 7 nm particle packings.

Hyperconfinement arises from a competition between confinement and incompressibility of the polymer. As polymer segments of fixed density are squeezed into narrower pores, they experience increased stretching and finally segregation. The crossover chain length where this segregation happens is dependent on the pore size and can be defined as follows.

Assuming a chain occupying a cylindrical pore, the polymer density occupying the pores is φ (= number of monomers/volume), the pore diameter is d, and the monomer size is a. The confined chain is perturbed in 2D and stretches out axially along the pore. At the onset of hyperconfinement, the chain is stretched out to the maximum extent along the pore length with monomers lined one after each other. For very narrow capillaries, this means that the chain fully occupies the pore, (end-end distance in melt)*(cross-sectional area of pores) = N*(monomer volume). This gives us $aN^{1/2}\frac{\pi d^2}{4} = Na^3$ at the critical crossover point N.

$$N \sim (d/a)^4 \quad (S9)$$

For 27 nm silica particles, the pore diameter is 9 nm and the critical crossover happens at around 500,000 g/mol. For the 7 nm silica particles, the pore diameter is 2.1 nm and the critical crossover occurs at around 1300 g/mol. Since these are order of magnitude estimation, the actual crossover chain weights likely are different from these estimates. Moreover, the geometry of the pores in nanoparticle packings is very different from that of cylindrical pores, which these estimations were based on. What is important to note is that the crossover happens at a much lower molecular weight for 7 nm silica particles than the 27 nm silica particles. Thus, hyperconfinement and chain segregation may play a role in the constant viscosity seen in PS confined in 7 nm silica nanoparticle packings.

## 4. Role of interfacial area of PS-silica on increase in $T_g$

Assuming that most of the NP surface is accessible to the polymer. The area/volume of the nanoparticles is given by $3/r$. The volume fraction of the randomly close-packed nanoparticles is 64% of the total volume. The rest of the pore space is filled by the polymers. Thus, over 36% of the film volume is occupied by the polymer with an interface between the polymer and the nanoparticles.

Total area $A = n * 4\pi r^2$ where r is the radius of the nanoparticle and n is the number of nanoparticles in the total film volume V. n is therefore given by $n = \frac{3V}{4\pi r^3}$. Thus, total interfacial area between polymer and NP goes as $\frac{3V}{r}$, and interfacial area between polymer and nanoparticle per unit volume is given by $\frac{3}{r}$.

Figure S3 shows the increase in $T_g$ seen with unentangled[10] and entangled PS confined in silica nanoparticle packings with the nanoparticle diameter varying from 7 to 77 nm against the inverse of particle diameter. There is an increase in $T_g$ with decreasing particle size but it does not correlate with the increased surface area. The data for 77 nm and 56 nm nanoparticles represent underconfined particles and we see that for the same surface area/volume the values of $\Delta T_g$ do not superimpose. With increasing confinement, we see that the values seem to be similar to each other at the same surface area/volume. However, there is a large gap between the values for PS-8k and 21k in 7 nm silica nanoparticle packings and PS-173k and above in 7 nm silica nanoparticle packings. Even accounting for the fact that the smaller molecular weight CaRI samples had cracks in them, the magnitude of $T_g$ increase for the entangled polymers is around 50 K more than unentangled polymers. The trends observed here indicate both increasing confinement and increased interfacial area contribute to the slowdown of segmental dynamics in the CaRI films.

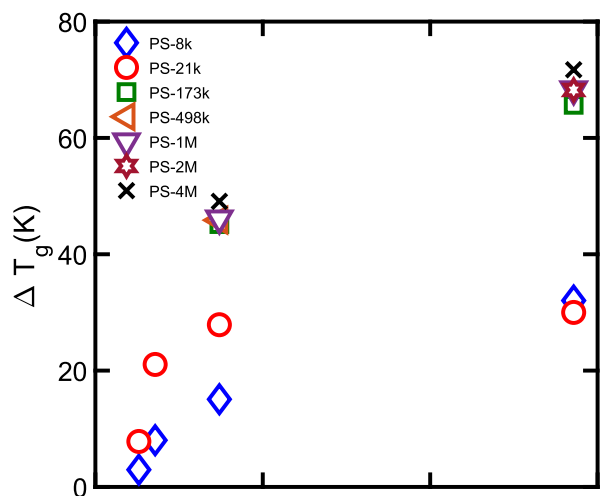

**Figure S3**: Increase in $T_g$ for confined unentangled(8k and 21k) and entangled(173k - 4 M g/mol) PS in silica nanoparticle packings of varying diameter(7-77 nm) plotted against inverse of particle diameter.

For entangled polymers, as the size of the nanoparticle decreases from 27 nm to 7 nm, $\Delta T_g$ increases by 45% wheas the surface area/volume increased by over 2.5 times. Increased confinement as well as an increase in the surface area between the nanoparticles and the polymer seem to be driving the increase in segmental slowdown in the smaller particle packings.

## 5. Viscosity data plotted against 1/r

A theoretical model developed by Yao et al for polymer flow in cylindrical nanopores has been used in the main text to explain two main effects in the capillary rise of PS into silica nanoparticle packings:

1) Slower-than-bulk rise of unentangled polymers – this increased viscosity is postulated to come from the buildup of a dead-zone of adsorbed polymers near the pore surface which reduces the effective pore diameter which provide additional viscous resistance to polymer flow. This results in an increase in viscosity which can be given by:

$$\frac{\eta}{\eta_0} = \left(\frac{R_{eff}}{R}\right)^{-4} \quad (S10)$$

where $R_{eff}$ is the reduced pore diameter due to the dead-zone, R is the original pore diameter and the confined and bulk viscosities are given by $\eta_0$ and $\eta$.

2) Faster-than-bulk rise of entangled polymers – this reduced viscosity arises from a reptation-like flow of polymers driven by the capillary pressure gradient in the nanoparticle packings. The expression for this is given by:

$$\frac{\eta}{\eta_o} = \frac{NR^2}{\varphi \eta_o} \quad (S11)$$

where N is the number of Kuhn segments in the chain, R is the pore radius, $\phi$ is a constant that contains the bulk polymer properties including the entanglement length and bulk viscosity is given by $\eta_0$.

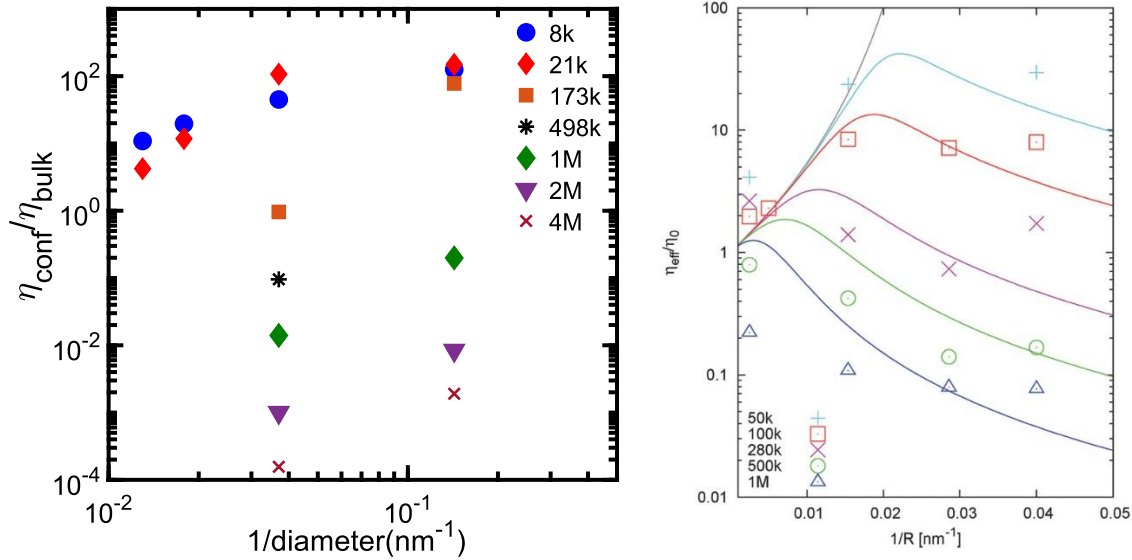

**Figure S4**: Normalized viscosity plotted against inverse of pore diameter for polystyrene in silica nanoparticle packings(left) and poly(ethylene-oxide) in Anodic Aluminium Oxide(AAO) pores(right). Reproduced with permission from references. [11]

Combining the two terms gives us:

$$\frac{\eta}{\eta_o} = \left[\left(\frac{R_{eff}}{R}\right)^4 + \frac{\varphi \eta_o}{NR^2}\right]^{-1} \quad (S12)$$

This equation can be used to describe the change in the effective viscosity of PS of any chain size in nanoparticle packings. We can apply equation S12 to our data in two ways: (a) keep R constant and adjust N; this has been done for entangled PS confined in 27 nm silica nanoparticles (Figure 5 of the main text). For entangled polymer of varying chain length and same average pore radius, we obtain $\eta \sim N^{0.8}$ which is similar to the prediction. (b) keep N constant and vary R – this is done as shown in Figure **S**4 where the abscissa has been chosen to be 1/R. Shown alongside is Figure S4 from Yao et al[11] and we can see the remarkable similarity in the trends even though the agreement between the theory and experimental results is imperfect.

# 6. Shear rate dependence in capillary rise – shear-induced thinning cannot explain the decrease in viscosity seen in the packings

The shear rate of the rising front of polymer in the silica nanoparticle packings is given by the following expression:

$$\dot{\epsilon} = \frac{1}{R_{pore}}\frac{dh}{dt} \qquad (S13)$$

where h is the height of the front at time t, r is the pore radius, and ἐ is the shear rate of the rising polymer. Since the data fits the Lucas Washburn Equation, we can use the model to obtain the gradient term as:

$$\frac{dh}{dt} = \frac{\gamma R_{pore} \cos\theta}{8\eta_{conf}\tau h} \qquad (S14)$$

where γ is the surface tension of PS, $R_{pore}$ is the pore radius(given here by 30% of the particle radius), θ is the contact angle at the polymer-air-silica interface, τ is the tortuosity of the packing, and $\eta_{conf}$ is the confined viscosity. This gives us:

$$\dot{\epsilon} = \frac{\gamma \cos\theta}{8\eta_{conf}\tau h} \qquad (S15)$$

Calculating this for the fastest rising polymer (PS-173kg/mol at 165 $^0$C in 27 nm Silica NP packings) gives us ἐ = 0.0099 s$^{-1}$. This can be compared to the critical shear rate at which shear thinning is observed in bulk PS melt.[12,13] Figure S5 shows the universal shear thinning behavior observed by polymer melts for PS – the normalized viscosity is plotted against Weissenberg number($Wi = \dot{\epsilon}\tau$, where τ is the relaxation time of the entire chain, given for PS-185k as 5.10 seconds).[13] For PS -173k, we can locate the corresponding shear rate on the abscissa as $Wi_m$ = 0.0495. This means that the normalized viscosity should be around 0.2. Thus, shear thinning only predicts an order of magnitude decrease in the viscosity. Thus, the decrease in the effective viscosity scaling with molecular weight that we observe cannot be explained solely by the shear thinning model.

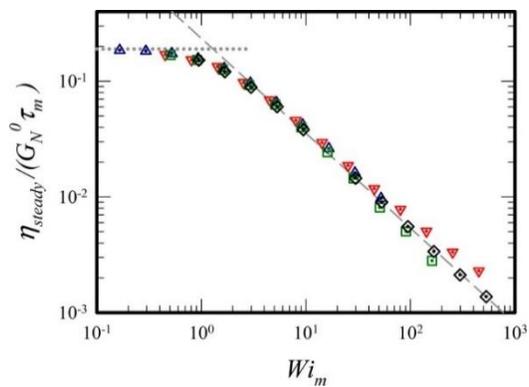

**Figure S5**: Shear-thinning of polystyrene PS133k (blue), PS285k/2k-47 (red), PS185k (green), and PS285k/2k-65 (black). The normalized viscosity is plotted against the Weissenberg number(which is the product of the terminal relaxation time and the shear rate). Adapted with permission from reference.[13]

## 7. Scaling with fixed distance from Tg

The confined viscosity is normalized by the bulk viscosity at the same temperature in the main text in Figure 5 despite the 50 K increase in $T_g$. $T_g$ of entangled polymers confined in the 25 nm silica nanoparticle packing are elevated with respect to bulk but they do not change significantly with molecular weight. This means that irrespective of our choice of temperature for measuring bulk viscosity, we are comparing the confined viscosity for different molecular weight polymer at the same distance from $T_g$. Thus, the choice of the temperature for the normalization should not matter. Here, we show this in Figure S6. Figure S6(left) shows the bulk viscosity values at measurement temperature (160 $^0$C) and 50 $^0$C above confined $T_g$ (~200 $^0$C). The scaling is done using the VFT equation parameters from the literature. Assuming that temp-time superposition relationships are valid, we can do a similar scaling of viscosity values for the confined polymer, shown also in Fig S6 (left). Then, the confined viscosity is normalized with the bulk at the two temperatures and a scaling exponent is obtained for viscosity with the number of segments N. The scaling stays independent of the choice of temperature at which the normalization is done.

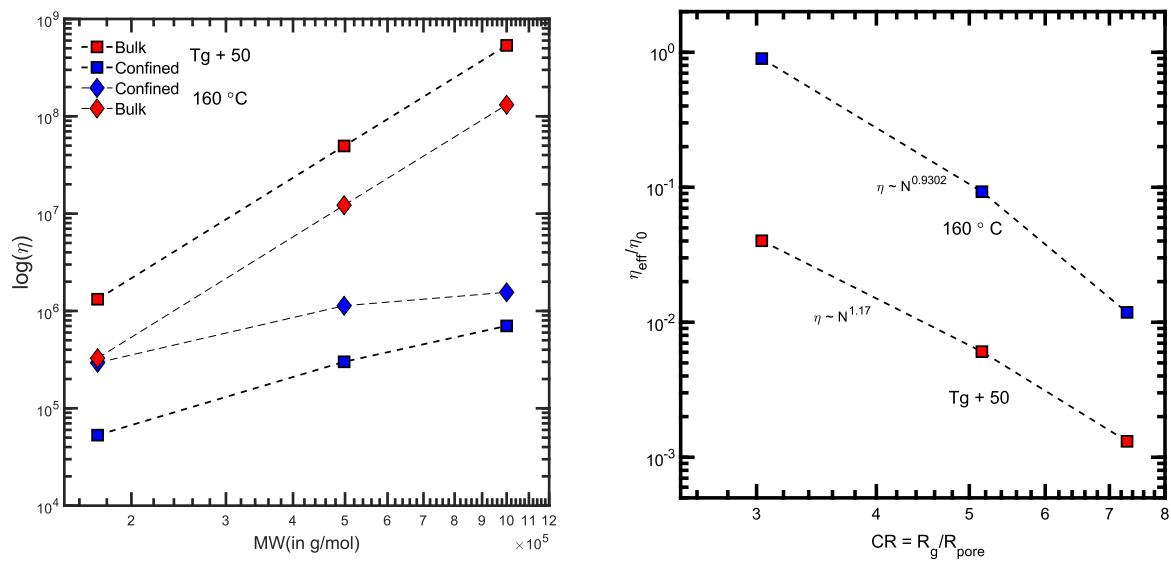

**Figure S6**: Confined(blue) and bulk(red) viscosity plotted of PS compared at the same temperature(diamonds) or at a fixed distance from Tg(squares). The ratio of the effective to the bulk viscosity is plotted against CR on the right and shows little change in the scaling of confined viscosity with number of segments in the polymer(N).